\documentclass{aa}  

\usepackage{array}
\usepackage{graphicx}
\usepackage{float}
\usepackage{academicons}
\usepackage{xcolor}
\usepackage{txfonts}
\usepackage[hidelinks]{hyperref}
\usepackage[normalem]{ulem} 

\begin{document}

    \title{Multiband neural network classification of ZTF light curves as LSST proxies}

   \author{T. Szklen\'ar 
          \inst{1,2} 
          \and
          A. B\'odi\inst{3}
          \and
          R. Szab\'o\inst{1,2,4,5}
          }

    \institute{Konkoly Observatory, HUN-REN Research Centre for Astronomy and Earth Sciences, Konkoly Thege Mikl\'os \'ut 15-17, H-1121, Budapest, Hungary, 
        \email{szklenar.tamas@csfk.org}
    \and
    HUN-REN CSFK, MTA Centre of Excellence, Konkoly Thege Mikl\'os \'ut 15-17,  H-1121, Budapest, Hungary
   \and
    Department of Astrophysical Sciences, Princeton University Peyton Hall, 4 Ivy Lane, Princeton, NJ 08544, USA 
   \and   
    ELTE E\"otv\"os Lor\'and University, Institute of Physics and Astronomy, Pázmány Péter sétány 1/a,  H-1117, Budapest, Hungary
    \and
    MTA–HUN-REN CSFK Lend\"ulet "Momentum" Stellar Pulsation Research Group, Konkoly Thege Mikl\'os \'ut 15-17,  H-1121, Budapest, Hungary
    }

   \date{Received May 09, 2025; accepted Dec 11, 2025}

 
  \abstract
    {Current and near-future sky survey programmes, like the Legacy Survey of Space and Time (LSST) of the Vera C. Rubin Observatory will produce vast amounts of data that will need new techniques to be developed to process them on reasonable timescales. Machine learning methods and properly trained neural networks proved to be efficient, fast, and reliable in performing a variety of  tasks such as classification of variable star light curves. Since LSST survey full sky data is not yet available (only the Data Preview 1 from various sky segments), we need to test our methods to be used on real LSST data on proxy data sets for now.}
    {In this project we use data obtained by Zwicky Transient Facility to develop and test a neural-network-based, multiband classification algorithm to classify periodic variable stars (i.e. pulsating variable stars and eclipsing binaries). The aim is to utilize the algorithm on LSST data once they become available.}
    {Phase-folded light curve images and period information were used from five different variable star types: Classical and Type II Cepheids, $\delta$ Scuti stars, eclipsing binaries, and RR Lyrae stars. The data is taken from the 17th  data release of ZTF, from which we used two passbands, g and r in this project. The periods were calculated from the raw data and this information was used as an additional numerical input in the neural network. For the training and testing process a supervised machine learning method was created, the neural network contains Convolutional Neural Networks concatenated with Fully Connected Layers.}
    {During the training-validation process the training accuracy reached 99\% and the validation accuracy peaked at 95.6\%. At the test classification phase three variable star types out of the 5 classes were classified with around 99\% of accuracy, the other two also had very high accuracy, 89.6\% and 93.6\%.}
   {Our results showed that utilizing phase-folded light-curves from multiple passbands and the periods as numerical data inputs we are able to train a neural network with outstanding accuracy.}


    \maketitle
%
\section{Introduction}

    As the first light of the Legacy Survey of Space and Time of the Vera C. Rubin Observatory [hereafter referred to as LSST] \citep{LSST} project is approaching, it is mandatory to develop unique methods to process the massive amount of data generated by this, as well as other, ongoing sky survey programmes. Among the countless challenges that can be pursued with the future LSST data, we embarked on the classification of (quasi)-periodic variable stars, exploiting the potential of neural networks.

    Using neural networks is becoming increasingly common, several methods have been published in various research fields. In astronomy and astrophysics it was used for transient candidate vetting in large-scale surveys \citep{gieseke2017}, studying the kinematics of cold gas in galaxies \citep{dawson2020}, classification of supernovae \citep{moller2020}, gravitational wave data  \citep{george2018}, exoplanet candidates \citep{osborn2020}, detecting solar-like oscillations \citep{hon2018}, exoplanets  \citep{dattilo2019}, as well as  oscillations in eclipsing binary light curves \citep{ulas2025} just to name a few use cases.
    
    Over the last few years, our team started to develop image-based classification methods of variable star light curves. 
    Space photometric missions (such as Kepler/K2 \citep{Kepler_Borucki_2016}, TESS \citep{TESS_2015}) provided quasi-continuous, densely sampled light curves. However, data from the Optical Gravitational Lensing Experiment [OGLE] \citep{Udalski2015} or the Zwicky Transient Facility [ZTF] \citep{Bellm_2019} are much more sparsely sampled, containing many large gaps. This difficulty can be overcome by using phase-folded light curves in the case of periodic variable stars (e.g. RR Lyrae stars, Cepheids, eclipsing binaries).
    The images of these phase-folded light curves serve as the main input for Convolutional Neural Networks \citep{CNN_basic_article} [CNN]. 
    
    Our first experiment \citep{Szklenaretal2020} showed that the image-based classification of periodic variable stars is reliable and fast [Paper I]. The results showed also the shortcomings of the method. Due to the similarity of the phase-folded light curves of certain variable types (i.e. degeneracies among Anomalous Cepheids and RR Lyrae stars) it was necessary to introduce additional numerical data of the physical parameters (e.g. period, magnitude) as inputs (Multiple-input Neural Network), which, in addition to a more accurate classification, allowed us, among other things, to identify the sub-types of variable stars with high accuracy  \citep{Szklenaretal2022} [Paper II]. 
    
    In this paper, we show how we  adapted our machine learning methods to be able to process periodic light curves coming from the ZTF sky-survey programme \citep{ZTF_Masci_2019}, which is the precursor of the ambitious LSST. Five variable star classes were used in this experiment: classical and Type II Cepheids \citep{OGLEIV_CCEP_T2CEP_ACEP_2017,OGLEIV_CEP_2018,OGLEIV_CEP_2020}, $\delta$ Scuti stars \citep{OGLEIV_DSCT_2020,OGLEIV_DSCT_2021}, eclipsing binaries \citep{OGLEIV_ECL_2016}, and RR Lyrae stars \citep{OGLEIV_RRLYR_2014,OGLEIV_RRLYR_2019}. Since OGLE provides well-categorized, reliably classified light curves, we cross-matched the overlapping sources between OGLE and ZTF, and as an intermediate step also with Gaia DR3 \citep{GaiaDR3_2023}. For the phase-folded light curves we used two passbands, $g$ and $r$, with the period of the variable stars serving as the numerical data input in the neural network. This way a multiband neural network was created, a method which was used before by other research groups as well, like \citep{MB1_Moreno-Cartagena_2025A&A...703A..41M, MB2_Chiong_2025arXiv250611637C, MB3_Becker2025A&A...694A.183B, MB4_Perez_2025MNRAS.540.3263P}
    
    This paper is structured as follows. In Section\,\ref{sec:data}, we present our data, how it was cross-matched with other databases and we also present the process of data augmentation. Section\,\ref{sec:neural_network} describes the structure of the neural network, while the results are discussed in Section\,\ref{sec:results_and_discussion}. The paper closes with our conclusions in  Section\,\ref{sec:conclusions}.

\section{Data}
\label{sec:data}

\begin{figure*}
   \centering
   \includegraphics[width=\textwidth]{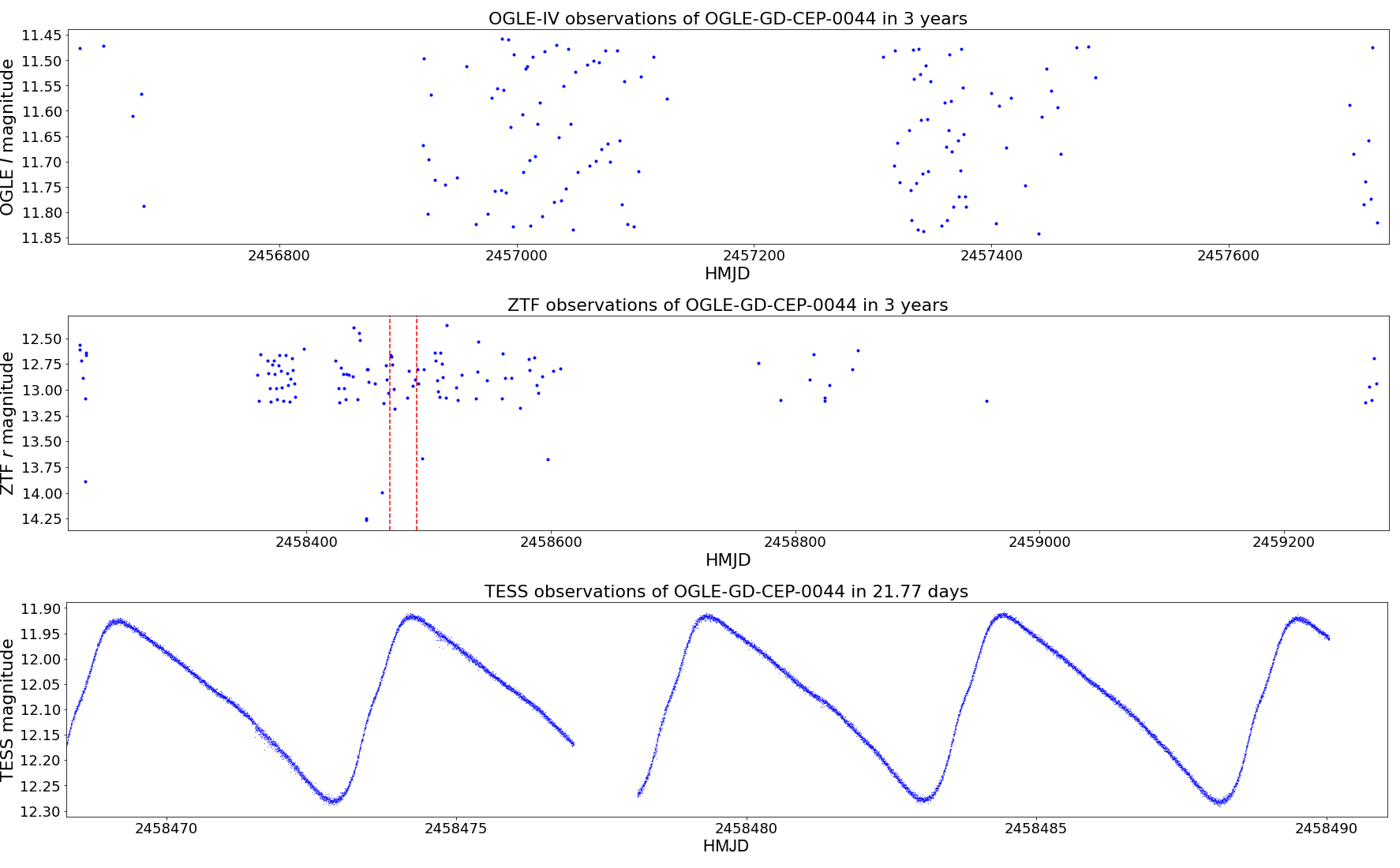}
     \caption{Visualization of the difference between sparse and dense sampling of photometric light curves. We chose a classical Cepheid  for this comparison. OGLE-GD-CEP-0044 is a relatively bright Cepheid with a period of 5.09 days. The upper graph shows OGLE-IV data for a 3 year long observation window, which has $130$ data points. The middle graph is observational data from the ZTF survey, also in a 3 year long observational window with $119$ data points. The third graph is a 21.77 days long observation from the TESS mission. The light curve contains 14\,898 measurements. The time span covered by the TESS observations are presented in the middle graph by two vertical dashed lines.}
     
    \label{fig:sparse_continous_data_comparison}
\end{figure*}

\begin{figure*}
   \centering
   \includegraphics[width=\textwidth]{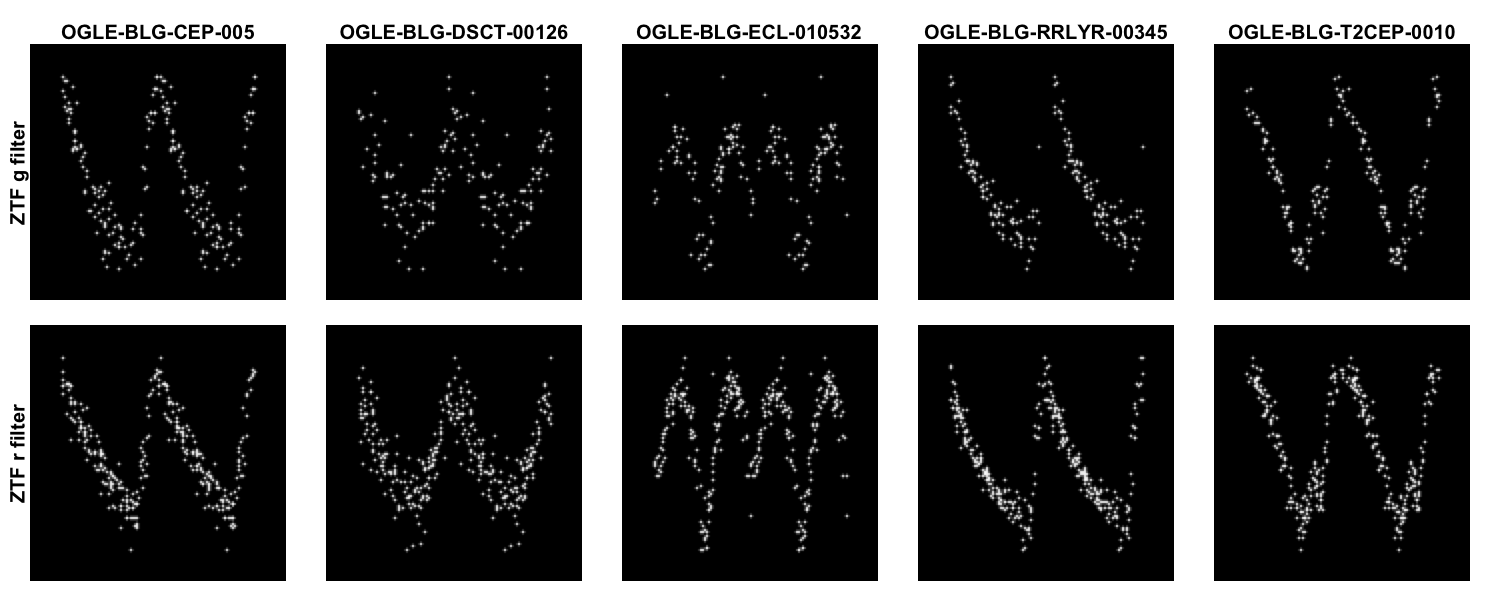}
     \caption{Example gallery of the phase-folded light curve images. The five columns represent the five main variable star types, which were used in this research: Classical Cepheids, $\delta$ Scuti stars, eclipsing binaries, RR Lyrae stars, and Type-II Cepheids. The top row shows g band data and the bottom row shows r band data of the given example of variable star type.}
    \label{fig:lightcurves_2passbands}
\end{figure*}

Ground-based sky-survey programmes often produce sparse data with gaps due to observational seasons, lunar phase, and diurnal variations. This presents many difficulties in data processing of periodic variable stars, as their light curve coverage is much sparser than data generated by dedicated space photometric telescopes, as it is presented in Figure\,\ref{fig:sparse_continous_data_comparison}. To get well-sampled light curves, the usual method in the case of periodic variable stars (pulsators, eclipsing binaries) is to phase-fold the light curves. This way we can reveal characteristic curve shapes that are specific to the different variable star types.

To simulate the processing and sampling of future LSST data, in our first experiments we used synthetic data from the  Extended LSST Astronomical Time-Series Classification Challenge (ELAsTiCC) \citep{ELAsTiCC_2023} dataset. However, the main purpose of this dataset was not to provide light curves for periodic variable stars, but to create and test end-to-end real-time pipelines for time-domain science in general , the ELAsTiCC simulated light curves still contain LSST-like light variations in six bands (ugrizY) for some variable star types, e.g. classical Cepheids or RR Lyrae stars. After some promising results, further analysis showed that while the light curve shapes and periods seem realistic, the generation of the light curves was not based on physical grounds, i.e. did not obey theoretical or observed trends and correlations among the observables. Hence, ELAsTiCC periodic variable star light curves were not adequate for our purposes (our training sets were always based on observed light curves, thus contain any astrophysics based correlations), we decided to switch to ZTF observations, as those are  very similar in structure to LSST. As ZTF is a pure sky-survey project, its database contains only observational data, but no cross-match between other catalogues is available for a given object. To be able to work with its data, first we had to cross-match ZTF objects with other catalogs to identify variable stars.\\
In this work we used the ZTF Data Release 17\footnote{ZTF DR17: \\https://irsa.ipac.caltech.edu/data/ZTF/docs/releases/dr17/ztf\_release\_notes\_dr17.pdf}. In the next subsection we explain how we handled the collected dataset.

\subsection{Cross-match between databases}
\label{sec:crossmatch}

The 51-inch (1.3m) Warsaw University Telescope used in the OGLE sky survey program is located at the Las Campanas Observatory, Chile. This survey collects information from 4 different sky areas, the Galactic Bulge (BLG), Galactic Disk (GD), the Large Magellanic Cloud (LMC) and the Small Magellanic Cloud (SMC). \\ In this work we used the OGLE-IV collection of variable stars and we only focused on the first two observational fields, the bulge and the Galactic Disk. 

The Zwicky Transient Facility (ZTF) at the Palomar Observatory uses the 48-inch (1.2m) telescope to scan the entire north sky in every two days in three passbands: $g$, $r$, and $i$. Although the telescope is located in the northern hemisphere, there is a sufficiently large overlap with the southern observatories, for example, with OGLE (above --28 degrees of declination). As our team has plenty experience with the latter, and we also used OGLE variable stars for training our machine learning algorithms (see Papers I and II), we decided to cross-match OGLE data with ZTF. It was obvious that this cross-match would not contain all pulsating variable stars from the OGLE catalogue, but it will be large enough to start work with. 

Five different major variable star types have been selected that already have sufficient data for this project. We knew that at least a couple of hundred unique stars are needed for a well sampled training dataset and we were aware that data augmentation and synthetic lightcurve generation is also needed. The number of selected variables per type and sub-type of star is presented in Table\,\ref{Tab:number_of_stars}.

\begin{table*}
\centering
\caption{Periodic variable stars and eclipsing binaries used in the article.}
\label{Tab:number_of_stars}
\begin{tabular}{|c|c|c|c|c|c|c|c|c|c|c|c}
\hline

\bf{Main class} & \multicolumn{6}{c|}{\bf{Field}} & \bf{Total data} & \bf{Total unique stars}\\
sub-class & \multicolumn{3}{c|}{\bf{BLG}} & \multicolumn{3}{c|}{\bf{GD}} & (overlapping fields) & \\
 & g & r & i & g & r & i & & \\

\hline
\hline
\bf{Classical Cepheid}     & \multicolumn{3}{c|}{\bf{167}}	& \multicolumn{3}{c|}{\bf{1\,840}}	& \bf{2\,007} & \bf{571} \\
\hline
1O	    & 12 & 16 & 4 & 233 & 219 & 85 & 569 & 154 \\
1O/2O     & 7 & 5 & 2 & 58 & 60 & 26 & 158 & 45 \\
1O/2O/3O	& 0 & 1 & 0 & - & - & - & 1 & 1 \\
F	        & 27 & 49 & 30 & 391 & 512 & 213 & 1\,222 & 353 \\
F/1O	    & 5 & 5 & 4 & 18 & 20 & 5 & 57 & 18 \\
\hline
\bf{$\delta$ Scuti}    & \multicolumn{3}{c|}{\bf{9\,970}}	& \multicolumn{3}{c|}{\bf{8\,315}}	& \bf{18\,285} & \bf{6\,882} \\
\hline
Multimode & 995 & 850 & 95 & 745 & 776 & 255 & 3\,716 & 1\,458 \\
Singlemode	& 3\,810 & 3\,657 & 563 & 2\,691 & 2\,797 & 1\,051 & 14\,569 & 5\,424 \\
\hline
\bf{Eclipsing binary}    & \multicolumn{3}{c|}{\bf{10\,216}}	& \multicolumn{3}{c|}{\bf{-}}	& \bf{10\,216} & \bf{5\,043} \\
\hline
C	    & 708 & 887 & 210 & - & - & - & 1\,805 & 854 \\
CV	    & 1 & 1 & - & - & - & - & 2 & 1\\
NC	    & 3\,172 & 4\,303 & 934 & - & - & -   & 8\,409 & 4\,188 \\
\hline
\bf{RR Lyrae}   & \multicolumn{3}{c|}{\bf{48\,497}}	& \multicolumn{3}{c|}{\bf{11\,092}}	& \bf{59\,589} & \bf{25\,702}\\
\hline
RRab	    & 15\,051 & 16\,128 & 3\,223 & 3\,360 & 3\,564 & 1\,406 & 42\,732 & 18\,317 \\
RRc	    & 6\,334 & 6\,480 & 992 & 1\,121 & 1\,136 & 381 & 16\,444 & 7\,223 \\
RRd	    & 119 & 115 & 19 & 45 & 49 & 17 & 364 & 140 \\
aRRd	   & 16 & 17 & 3 & 6 & 5 & 2 & 49 & 22 \\
\hline
\bf{Type II Cepheid}    & \multicolumn{3}{c|}{\bf{1\,018}}	& \multicolumn{3}{c|}{\bf{357}}	& \bf{1\,375} & \bf{568} \\
\hline
1O   & 2 & 2 & 0 & 0 & 1 & 0 & 5 & 2 \\
BLHer   & 203 & 235 & 51 & 53 & 60 & 20 & 622 & 260 \\
RVTau     & 68 & 69 & 23 & 31 & 31 & 12 & 234 & 97 \\
WVir   & 142 & 160 & 42 & 48 & 56 & 15 & 463 & 192 \\
pWVir   & 7 & 10 & 4 & 11 & 12 & 7 & 51 & 17 \\
\hline
\end{tabular}
\tablefoot{
The number of variable stars (both main and sub-classes) collected from the OGLE-IV database which overlap with the ZTF observational fields. The two fields are the Galactic Bulge (BLG) and the Galactic Disc (GD).  The subclass names are the same as they are presented in the OGLE-IV catalog. Many of the stars in ZTF that are observed in multiple overlapping fields have different identifiers in each field. Additionally, the stars receive different id in every filter ($g$, $r$, $i$) as well. These are the reasons why there are more observational data than the real number of unique stars. 
}
\end{table*}

As \citet{Wilson_Naylor_crossmatch} 
and others pointed it out on numerous occasions, LSST will have such crowded fields that the standard cross-matching algorithms will fail. For us, who would like to gather information from multiple catalogues, source identification is crucial. With the help and advice of Wilson \& Naylor (priv. commmunication), we started to cross-match pulsating variable stars and eclipsing binaries. As the LSST point sources in the brightness range between LSST’s saturation limit and the Gaia limiting magnitude will be cross-matched with the Gaia catalog \footnote{Science Validation of LSST Data Release Processing - http://pstn-039.lsst.io}, the Gaia DR3 was used as a starting point for comparing the other two databases. First we cross-matched the OGLE-IV catalog with Gaia, after that we compared Gaia with ZTF data using a probalistic approach. The last step was to merge the results and create the final dataset.

The details of the cross-matching side project will be published in a forthcoming paper, since it is outside the scope of this work. In the following, we only give a brief overview of the process.

Fortunately these three instruments have very similar light gathering capabilities and limiting magnitudes, therefore the crowding problem is less severe when comparing them to each other. However, we had to take into account the various passbands of the surveys. Every star from OGLE is measured in the Johnson-Cousins UBVRI photometric system \citep{UBVRI_passbands}, mainly in I filter, but the majority of the star have V measurements as well. Gaia is using its unique G passband, which is operated over a much wider wavelength range, but in numerous cases the narrower Gaia BP and Gaia RP magnitudes are also available. Finally, the instrument of ZTF is equipped with yet another filter set, namely SDSS-like $g$, $r$, and $i$ \citep{SLOAN_1996}.

For efficient Gaia queries, the entire Gaia \texttt{source} and \texttt{var\_summary} tables were downloaded and a database was created from these on the HUN-REN Cloud \citep{HUN-REN_Cloud_H_der_2022}. The database is based on the latest version of MySQL \citep{MySQL}. 
During the cross-match process, the source's position and brightness were taken into account, and the closest neighbour method was used with additional matching parameters. The possible matches were categorised by the separation and the similarity of magnitude information. For the latter we compared the OGLE I passband with Gaia RP mean magnitudes (if it was available). Gaia RP covers similar, but wider wavelength range as the I passband. Despite the similarities, we used 1.0 magnitude difference limit for the comparison. According to preliminary statistics, the OGLE-IV have very precise coordinates. The average separation between OGLE and Gaia is less than 0.5 arcseconds, so 1 arcsecond was used as the limit of separation. \\
The matches were put into 4 categories: the source received score 3 if both the separation and the brightness difference are lower than the limit. Score 2 meant that the star is within the separation limit, Gaia RP magnitude is available, but the difference of the magnitudes is greater than the threshold. Those sources received score 1, where Gaia RP magnitudes were not available, but the separation of the stars are within the set limit. Those OGLE stars which can not be found in the Gaia DR3 catalog received score 0. Usually these stars are very faint, for example $\delta$ Scuti stars. \\
In the vast majority of cases only one source matched in a given field, in other cases -- where we had to deal with more than one candidate sources -- the closer separation won. A few sources in question had to be reviewed by eye. This way we got a very well matched list of stars. \\
Stars in the ZTF sky survey are observed in $g$, $r$, and $i$ filters. They receive a different \texttt{objectid} in each filter and in each observational field. For this reason they were treated as separate objects, the cross-match process was run for all filters separately. Because of the $i$ filter data are mostly available only for private projects, are not public and the available observations are very sparse, we decided the use only the $g$ and $r$ filters. Our team does not work using the various ZTF alert systems, where a given variable source receives a single ID (“alert id”). We used the ZTF17 database, where a different (“ztf objectid”) is used. Unfortunately, these identifiers are not connected. One can find out them only by doing an additional cross-match, which is obviously time-consuming and not practical. We bulk downloaded the entire ZTF DR17 database instead and searched and filtered out our sources. 

For the cross-matching we used a similar method as before. The sky survey's website \footnote{ZTF catalog search: https://irsa.ipac.caltech.edu/Missions/ztf.html} offers a multiple-object query, which outputs a list of objects based on the nearest neighbour method. Because of the wider cone radius of the search -- 5 arcseconds --, this contains a lot of inappropriate data and therefore needs further filtering, but it is very useful in the sense that we don't have to query the whole database. Unfortunately, the SDSS $g$ and $r$ filters do not resemble neither the Gaia BP or RP magnitudes nor the OGLE-IV Johnson-Cousins I or V filters, so comparing the brightnesses was done with a slightly larger margin of error, 1.5 magnitudes.

\section{Preprocessing of the data}
\label{sec:dataselection}

\subsection{Data selection, data augmentation}
\label{sec:dataselection}
The main goal of our team is to create methods for classification of periodic variable stars in the LSST database. After the cross-match process was finished, we used the collected ZTF data as LSST proxy data. As it is still uncertain that the periods will be calculated in the alert systems of LSST, we decided to do this job in this project ourselves. The period was calculated for every star in every filter. With this information we were able to phase-fold the raw observational $g$ and $r$ band observational data. These, in most cases, quite distinctive light curves with the additional period information can be separated into known variable star types. The phase-folded light curves were stored as 128x128 pixel, 1-bit images, with black background and white plotted dots. For additional information on the procedure of creating the images of the light curves we refer to Paper I.

Data augmentation \citep{Shorten2019} was necessary because of the order of magnitude difference in the number of representatives falling into the given variable star classes. On the one hand – as it can be seen in Table \ref{Tab:number_of_stars} –, due to the limited overlap of the observed sky areas, there are, for example, fewer Classical (571 unique sources) and Type II Cepheids (568 unique sources), than $\delta$ Scuti (6882 unique sources) or RR Lyrae stars (25702 unique sources). On the other hand, there are significant differences in the intrinsic occurrence rates of certain variable types, as well as in their detection efficiencies. This number of difference, the so called data augmentation problem, made the generation of synthetic data crucial. Synthetic phase-folded light curves were generated in the same way as we described in Paper II. After the process we had 4\,500 light curves both in $g$ and $r$ filters from five different variable star types: classical Cepheids, $\delta$ Scutis, eclipsing binaries, RR Lyrae stars and Type II Cepheids, altogether 45\,000 light curves.

\subsection{Single band period calculation}
\label{sec:dataselection}
The periods were calculated for every variable star in this sample, both for $g$ and $r$ filters using the $MultiHarmonicFitter$ function of the \texttt{seismolab} software package \citep{seismolab_Bodi_2024}.  This  analysis was particularly important for us because it allowed to test how reliable we would be able to calculate the period of variable stars for this sparse data in the first few years of the upcoming LSST survey. As we will have very few data points per year (70-80 on average) and per filter (10-15 per filter annually), we could test how our methodology evolves from year to year. Therefore, we calculated the periods for 1-, 2-, 3-, and 5-year long data segments per variable star type and compared the results to the known periods from the OGLE-IV database. \\
During this calculation process our first criterion was the number of data points collected in the given observational window. Based on preliminary tests and our expertise, 20 data points were defined as the minimum required to be able to calculate periods for most of the variable star types. Because of the different period information and the poorly sampled data it was obvious that for some stars we will need much longer observational time base. Every star (both in $g$ and $r$ filters) were flagged if the first criterion was not met. \\
The second criterion was a comparison of the calculated period with the 'ground-truth' data from the OGLE-IV catalogue. For those stars where plenty of measurements were collected (first criteria met), but the period calculation were not precise enough and exceeded 15 percent difference compared to the OGLE-IV periods, the frequency calculation boundaries were fine-tuned based on the 'ground-truth' data. The \texttt{MultiHarmonicFitter} function's $minimum\_frequency$ and $maximum\_frequency$ parameters were changed and the periods were searched within these new intervals.
In this way, we obtained a success rate for the period calculation. The aggregated result can be seen in Table\,\ref{Tab:period_calculation}. Based on these results, we find that we will have reliable period information for the mentioned variable star types (except for the longest period eclipsing binaries) after three years of continuous LSST operation.

\subsection{Multiband period calculation}
\label{sec:multiband_period}

In the case of variable star data collected by ground-based observations similar to OGLE, ZTF, and especially LSST, which have very poor sampling, finding and calculating the correct period is a particularly difficult task. Therefore, instead of the methods used previously, new algorithms are needed that make better use of the data collected by the LSST in six different filters and allow for so-called multiband period searches (e.g.,  \citealt{multiband_periodigrams_VanderPlas_Ivezic_2015}) when using them together.\\
Similar to other research groups, for example \citet{light_curve_recovery_Criscienzo_2023} or \cite{Pablo_2018}, our team is interested in using data collected with multiple filters to calculate periods, thereby somewhat offsetting the problems arising from poor sampling. However, the main topic of our paper was not to test or develop methods for calculating periods, partly because these were already available in the OGLE-IV database, calculating periods using single filters or multiband methods is important, so we also ran the latter for different time windows (1, 2, 3, and 5 years). 
For the multiband period calculation the \texttt{LombScargleMultiband} class was used from the Astropy \citep{Astropy_2022ApJ...935..167A} package. After multiple runs we decided to use $20$ points for the \texttt{samples\_per\_peak} and $25$ for the \texttt{maximum\_frequency} parameters.\\ 
For comparison, the multiband method performed better for shorter time windows (1 and 2 years). The results are significantly better for the Classical Cepheids and the Type-II Cepheids, which have longer (couple of days) periods. The algorithm performed worse for the $\delta$ Scuti stars and eclipsing binaries, which have very short periods, in most cases less than a fraction of a day. In the case of the eclipsing binaries, we not only checked whether the period was correct, but also whether the result obtained was half or twice the original period. We accepted these values and those that differed slightly (maximum 5\%) from the baseline value.\\
The aggregated result can be seen in Table\,\ref{Tab:multiband_period_calculation}. Based on these results, we find that multiband period calculations have much potential if the observational window is long enough to have sufficient number of observations in every filter.

\begin{table*}
\scriptsize
\centering
\caption{Success rate of period calculations using single band data.}

\label{Tab:period_calculation}
\setlength\tabcolsep{3pt}
\setlength\extrarowheight{5pt}
\begin{tabular}{c||c|c|c|c|c|c||c|c|c|c|c|c||c|c|c|c|c|c||c|c|c|c|c|c}

\hline
\bf{Variable star type} & \multicolumn{6}{c||}{\bf{1 year}} & \multicolumn{6}{c||}{\bf{2 years}} & \multicolumn{6}{c||}{\bf{3 years}} & \multicolumn{6}{c}{\bf{5 years}}\\
\hline
\hline
& \multicolumn{3}{c|}{$g$ filter} & \multicolumn{3}{c||}{$r$ filter} & \multicolumn{3}{c|}{$g$ filter} & \multicolumn{3}{c||}{$r$ filter} & \multicolumn{3}{c|}{$g$ filter} & \multicolumn{3}{c||}{$r$ filter} & \multicolumn{3}{c|}{$g$ filter} & \multicolumn{3}{c}{$r$ filter}\\
\hline

\bf{Classical Cepheids} &
\textcolor{red}{$50.9$} & \textcolor{teal}{$40.3$} & \textcolor{orange}{$8.8$} & 
\textcolor{red}{$50.8$} & \textcolor{teal}{$40.0$} & \textcolor{orange}{$9.1$} & 
\textcolor{red}{$41.0$} & \textcolor{teal}{$51.3$} & \textcolor{orange}{$7.7$} & 
\textcolor{red}{$40.5$} & \textcolor{teal}{$51.5$} & \textcolor{orange}{$8.0$} & 
\textcolor{red}{$32.4$} & \textcolor{teal}{$59.8$} & \textcolor{orange}{$7.9$} &
\textcolor{red}{$26.0$} & \textcolor{teal}{$63.0$} & \textcolor{orange}{$10.9$} &
\textcolor{red}{$25.8$} & \textcolor{teal}{$63.4$} & \textcolor{orange}{$10.8$} &
\textcolor{red}{$19.2$} & \textcolor{teal}{$69.1$} & \textcolor{orange}{$11.7$} \\ 

\bf{Delta Scuti stars} & 
\textcolor{red}{$65.1$} & \textcolor{teal}{$33.7$} & \textcolor{orange}{$1.1$} & 
\textcolor{red}{$63.1$} & \textcolor{teal}{$35.1$} & \textcolor{orange}{$1.7$} & 
\textcolor{red}{$36.9$} & \textcolor{teal}{$60.7$} & \textcolor{orange}{$2.4$} & 
\textcolor{red}{$18.5$} & \textcolor{teal}{$78.5$} & \textcolor{orange}{$3.1$} & 
\textcolor{red}{$21.7$} & \textcolor{teal}{$74.6$} & \textcolor{orange}{$3.7$} &
\textcolor{red}{$14.8$} & \textcolor{teal}{$82.1$} & \textcolor{orange}{$3.1$} &
\textcolor{red}{$16.1$} & \textcolor{teal}{$81.4$} & \textcolor{orange}{$2.5$} &
\textcolor{red}{$13.5$} & \textcolor{teal}{$83.3$} & \textcolor{orange}{$3.2$} \\ 

\bf{Eclipsing binaries} & 
\textcolor{red}{$100.0$} & \textcolor{teal}{$0.0$} & \textcolor{orange}{$0.0$} & 
\textcolor{red}{$82.2$} & \textcolor{teal}{$1.6$} & \textcolor{orange}{$16.3$} & 
\textcolor{red}{$71.1$} & \textcolor{teal}{$1.6$} & \textcolor{orange}{$27.3$} & 
\textcolor{red}{$4.4$} & \textcolor{teal}{$9.1$} & \textcolor{orange}{$86.5$} & 
\textcolor{red}{$46.3$} & \textcolor{teal}{$3.5$} & \textcolor{orange}{$50.2$} &
\textcolor{red}{$3.9$} & \textcolor{teal}{$9.0$} & \textcolor{orange}{$87.1$} &
\textcolor{red}{$33.9$} & \textcolor{teal}{$4.2$} & \textcolor{orange}{$61.9$} &
\textcolor{red}{$3.2$} & \textcolor{teal}{$8.3$} & \textcolor{orange}{$88.5$} \\ 

\bf{RR Lyrae stars} & 
\textcolor{red}{$72.4$} & \textcolor{teal}{$20.1$} & \textcolor{orange}{$7.6$} & 
\textcolor{red}{$72.9$} & \textcolor{teal}{$17.0$} & \textcolor{orange}{$10.0$} & 
\textcolor{red}{$34.0$} & \textcolor{teal}{$48.2$} & \textcolor{orange}{$17.8$} & 
\textcolor{red}{$11.7$} & \textcolor{teal}{$66.2$} & \textcolor{orange}{$22.1$} & 
\textcolor{red}{$19.5$} & \textcolor{teal}{$60.1$} & \textcolor{orange}{$20.5$} &
\textcolor{red}{$10.0$} & \textcolor{teal}{$71.2$} & \textcolor{orange}{$18.8$} &
\textcolor{red}{$14.3$} & \textcolor{teal}{$71.5$} & \textcolor{orange}{$14.3$} &
\textcolor{red}{$8.9$} & \textcolor{teal}{$76.4$} & \textcolor{orange}{$14.6$} \\ 

\bf{Type-II Cepheids} & 
\textcolor{red}{$73.1$} & \textcolor{teal}{$19.1$} & \textcolor{orange}{$7.8$} &
\textcolor{red}{$77.4$} & \textcolor{teal}{$17.0$} & \textcolor{orange}{$5.7$} & 
\textcolor{red}{$36.8$} & \textcolor{teal}{$43.2$} & \textcolor{orange}{$20.0$} & 
\textcolor{red}{$18.2$} & \textcolor{teal}{$59.4$} & \textcolor{orange}{$22.3$} & 
\textcolor{red}{$23.7$} & \textcolor{teal}{$53.5$} & \textcolor{orange}{$22.8$} &
\textcolor{red}{$16.0$} & \textcolor{teal}{$64.8$} & \textcolor{orange}{$19.2$} &
\textcolor{red}{$18.6$} & \textcolor{teal}{$66.5$} & \textcolor{orange}{$14.9$} &
\textcolor{red}{$14.2$} & \textcolor{teal}{$71.5$} & \textcolor{orange}{$14.3$} \\ 

\hline
\end{tabular}
\tablefoot{Success rate (in percentage, rounded to 1 decimal place) of period calculations using ZTF data with different observation time spans in both $g$ and $r$ filters. The calculations were compared to the OGLE-IV periods.
The red colours mean unsuccessful calculations due to insufficient data points (less than 20), the teal coloured percentages mean that the period calculations were successful. The orange coloured text shows those stars, which period needed additional fine-tuning, based on the OGLE-IV data.
} 
\end{table*}

\begin{table*}
\scriptsize
\centering
\caption{Success rate of period calculations using the multiband Lomb-Scargle method.}

\label{Tab:multiband_period_calculation}
\setlength\tabcolsep{10pt}
\setlength\extrarowheight{5pt}
\begin{tabular}{c||c|c||c|c||c|c||c|c}

\hline
\bf{Variable star type} & \multicolumn{2}{c||}{\bf{1 year}} & \multicolumn{2}{c||}{\bf{2 years}} & \multicolumn{2}{c||}{\bf{3 years}} & \multicolumn{2}{c}{\bf{5 years}}\\
\hline
\hline

\bf{Classical Cepheids} &
\textcolor{red}{$31.3$} & \textcolor{teal}{$68.7$} &
\textcolor{red}{$23.1$} & \textcolor{teal}{$76.9$} &
\textcolor{red}{$16.9$} & \textcolor{teal}{$83.1$} &
\textcolor{red}{$14.4$} & \textcolor{teal}{$85.6$} \\ 

\bf{Delta Scuti stars} & 
\textcolor{red}{$87.0$} & \textcolor{teal}{$13.0$} &
\textcolor{red}{$67.7$} & \textcolor{teal}{$32.3$} &
\textcolor{red}{$61.9$} & \textcolor{teal}{$38.1$} &
\textcolor{red}{$53.1$} & \textcolor{teal}{$46.9$} \\

\bf{Eclipsing binaries} & 
\textcolor{red}{$91.0$} & \textcolor{teal}{$9.0$} &
\textcolor{red}{$66.0$} & \textcolor{teal}{$34.0$} &
\textcolor{red}{$62.0$} & \textcolor{teal}{$38.0$} &
\textcolor{red}{$54.0$} & \textcolor{teal}{$46.0$} \\

\bf{RR Lyrae stars} & 
\textcolor{red}{$69.0$} & \textcolor{teal}{$31.0$} &
\textcolor{red}{$34.8$} & \textcolor{teal}{$65.2$} &
\textcolor{red}{$28.1$} & \textcolor{teal}{$71.9$} &
\textcolor{red}{$15.7$} & \textcolor{teal}{$84.3$} \\ 

\bf{Type-II Cepheids} & 
\textcolor{red}{$43.3$} & \textcolor{teal}{$56.7$} &
\textcolor{red}{$22.2$} & \textcolor{teal}{$77.8$} &
\textcolor{red}{$18.3$} & \textcolor{teal}{$81.7$} &
\textcolor{red}{$14.1$} & \textcolor{teal}{$85.9$} \\ 

\hline
\end{tabular}
\tablefoot{Success rate (in percentage, rounded to 1 decimal place) of period calculations using the multiband Lomb-Scargle method with ZTF data. Similar to the previous calculation, we used different time windows. Only two filters ($g$ and $r$) were present in the multiband calculation and these were compared to the OGLE-IV periods.
The red colours mean unsuccessful calculations, the teal coloured percentages mean that the period calculations were successful.}
\end{table*}

\subsection{Training, validation and test dataset}
\label{sec:training}

During the pre-processing phase, the dataset mentioned in the previous subsection was divided into two, non-overlapping parts, one for training and validation (4\,000 stars) and one for testing purposes (500 stars). Later, the training-validation dataset was again separated into two parts with a {0.7 ratio: 70\% was used for training and 30\% was used for validation. We made sure that all variable star types are equally represented in the training and validation samples with either original or augmented light curves.

\subsection{Variable star classes and labels}
\label{sec:classes}

A supervised neural network method was used during the training-validation phase. This type of neural network needs a well labelled dataset, where every category has a unique identification. In our previous work, we were able to train the network using not only the main variable star types, but also the sub-types, with outstanding accuracy (Paper II). However, in this work we could use  samples only from the OGLE-IV galactic disc (GD) and galactic bulge (BLG) fields, since neither the LMC nor the SMC is visible in ZTF. Because of this, some pulsating variable star types that were mentioned in our previous papers} -- for example Anomalous Cepheids -- are missing from the current dataset. Moreover, some variable star classes had unfortunately just a few targets in the overlapping catalogues. If we had broken them down into further groups, we would have had even less data to work with. This is why we decided  train  only the main variable star types - Classical Cepheids, $\delta$ Scuti stars, eclipsing binaries, RR Lyrae stars and Type II Cepheids - in this experiment.

\section{Neural Network}
\label{sec:neural_network}

\begin{figure*}[h!]
   \centering
   \includegraphics[width=\textwidth]{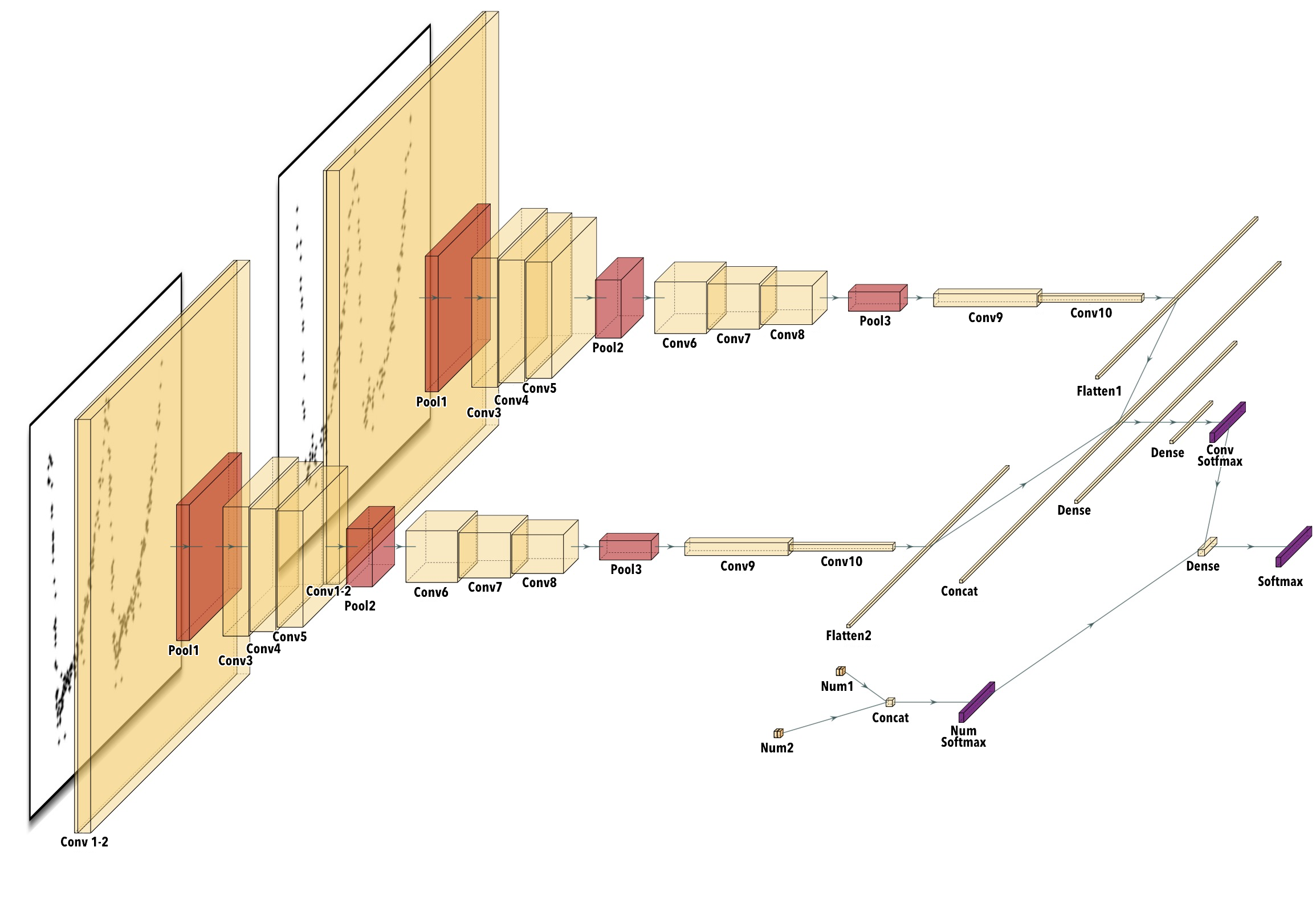}
     \caption{Architecture of our neural network which can classify image-based, phase-folded light curves in multiple (in this case two) passbands with additional numerical input (in this case the periods) measured in each passband. From left to right: two light curves for the same variable star in two different passbands, $g$ and $r$, are the image inputs of the neural network. These images are processed by two identical Convolutional Neural Networks (CNN) and then are concatenated together to make classification based only on the image (light curve) information. 
     The additional, numerical inputs are the passband-dependent periods of the given variable star, which also processed in a much simpler, Fully-connected Neural Network.
     These two inputs and their results are concatenated in the end to make the final classification result. }
     
    \label{fig:neural_network_architecture}
\end{figure*}

The architecture of the neural network contains two major parts (see Figure\,\ref{fig:neural_network_architecture} and Table\,\ref{Tab:hyperparameters}). The first part is a Convolutional Neural Network (aka CNN). The $g$- and $r$-filter phase-folded light curves are the image-based inputs for two separate CNNs, which are concatenated in the end. The second part of the network handles the numerical input, in this case the periods. Both parts perform the classification independently, making estimates of the variable star's class. Both are concatenated in the end and a final classification is made for the type.

This paper is not intended to describe the different types of
neural network layers, only the basic structure is described below. Interested readers are encouraged to consult our Paper I., which contains such descriptions. The network was developed using the Keras API built over TensorFlow \citep{tensorflow}, an open source platform for machine learning. Every code was written in Python.    

\subsection{Light curve (image) classification}
\label{sec:CNNs}

The classification of the image-based data is performed by two identical CNNs (as we described in the previous subsection), which can be separated into four different parts. The first part consists of two convolutional layers with $7\times$7 and $5\times$5 convolutional windows, followed by a dropout and a pooling layer. The purposes of this first section is the extraction of low-level features and the resizing of input images.

The high-level features are extracted in the following parts, separated into three blocks. The first two contains the same architecture, having three convolutional layers using $3\times$3 convolutional windows, ending with one dropout and one pooling layer. The last block contains two convolutional layers with the usual dropout and pooling layers. After this we apply a flatten layer to create a usable input for the fully connected and the softmax layers.

\begin{table}
\footnotesize
\centering
\caption{Hyperparameters of the neural network} 
\label{Tab:hyperparameters}
\begin{tabular}{rc}
\hline
Parameter & Value \\
\hline
\hline

\multicolumn{2}{c}{\emph{Architecture}} \\ \\

Starting convolution window & [$7\times7]$ \\
Convolution stride & 1\\
Convolution padding & 0\\
Convolution activation & ReLU \\
Dropout probability & 0.3 \\
Pooling type & MaxPooling \\
Pooling size & $2\times2$ \\
Number of convolution layers & 10 \\
Number of pooling layers & 4\\
Number of dense layers & 6\\
Dense activation function & ReLU\\ 

\hline
\multicolumn{2}{c}{\emph{Optimization}} \\ \\

Batch size & 32 \\
Learning rate & $2.5\times10^{-4}$\\
Optimizer & Adam \\
Loss function & Categorical crossentropy\\
Early stopping $\Delta$ and patience & $10^{-4}$, 19 epochs \\
\hline
\end{tabular}
\end{table}

\subsection{Numerical data}
\label{sec:numerical_data}

The period of the variable stars served as numerical inputs -- basically floating point numbers -- in the network. Periods calculated from the $g$ and $r$ band light curves were handled by two fully-connected layers with a $softmax$ activation function. The output of this layer is the classification based on periods only.

\subsection{Concatenating the outputs}
\label{sec:concat}

As both the image and the numerical data have the same number of output classes, we can concatenate them and use the result as a new input for further joint classification. The two (or more) concatenated outputs are fed as a single input to a fully-connected layer with 64 units, the output of which is sent to a $softmax$ dense layer which performs the final classification.

\subsection{Training optimization, early stopping} 
\label{sec:optimizers}

In our recent works we performed thorough tests for the most optimal hyperparameters. The Adam (Adaptive Moment estimation) optimizer with a learning rate of $0.00025$ was used similar to our previous papers. The batch value of 32 were used during the training and validation process. This value controls the number of training examples in one forward/backward pass, lower values require less memory and these so-called mini batches typically train the network faster. With this parameter it took $405$ iterations to complete $1$ training epoch. Table\,\ref{Tab:hyperparameters} lists the hyperparameters of the neural network.

To avoid overfitting, we applied an \texttt{EarlyStopping} callback during training, which monitors the change of the validation loss value. We set the \texttt{min\_delta} and \texttt{patience} values to $10^{-4}$ and 19, respectively, which values were chosen during hyperparameter optimization.

\subsection{Randomized phase test}
\label{sec:random_phase_test}

At the referee's request, we performed a test to randomize the phase. The suggestion was that in the case of variable stars, one should not introduce external information, e.g. by ensuring that the maximum of the pulsating variable stars or the minimum of the eclipsing binaries fell at the zero phase, which is the general procedure in period determination methods. During the test, we randomized the phase of the maximum/minimum value in the 0-1 interval, thereby filtering out this bias. This involved regenerating all light curves and rerunning the training and testing algorithms. According to our test, we could reproduce the results that were obtained without the phase-randomizing step, which means that our algorithm does not learn the phase information, but rather works based on the light curve shape.
We are attaching the confusion matrix in the Appendix (see Figure\,\ref{fig:appendix_randomphase_cm}).

\section{Results}
\label{sec:results_and_discussion}

In the following, we evaluate the training performance of the neural network and present our classification results on the test sample of
the variable stars. 

\subsection{Training performance}
\label{sec:training_performance}

\begin{figure}
   \centering
   \includegraphics[width=\columnwidth]{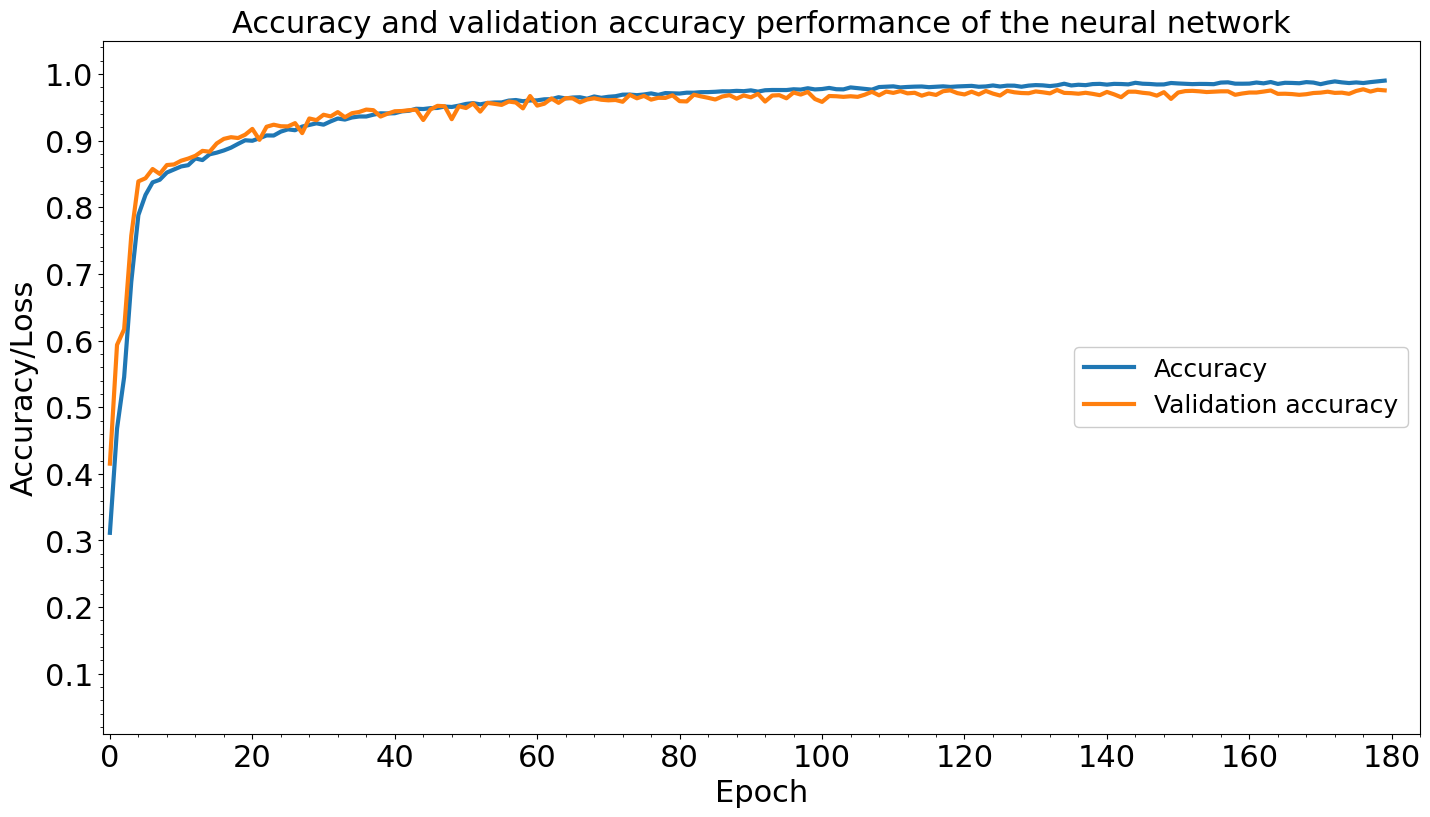}
     \caption{History curves of the neural network show how the
    training and validation accuracy evolved during the training and validation phases.
    }
    
    \label{fig:accuracy_performance}
\end{figure}

\begin{figure}
   \centering
   \includegraphics[width=\columnwidth]{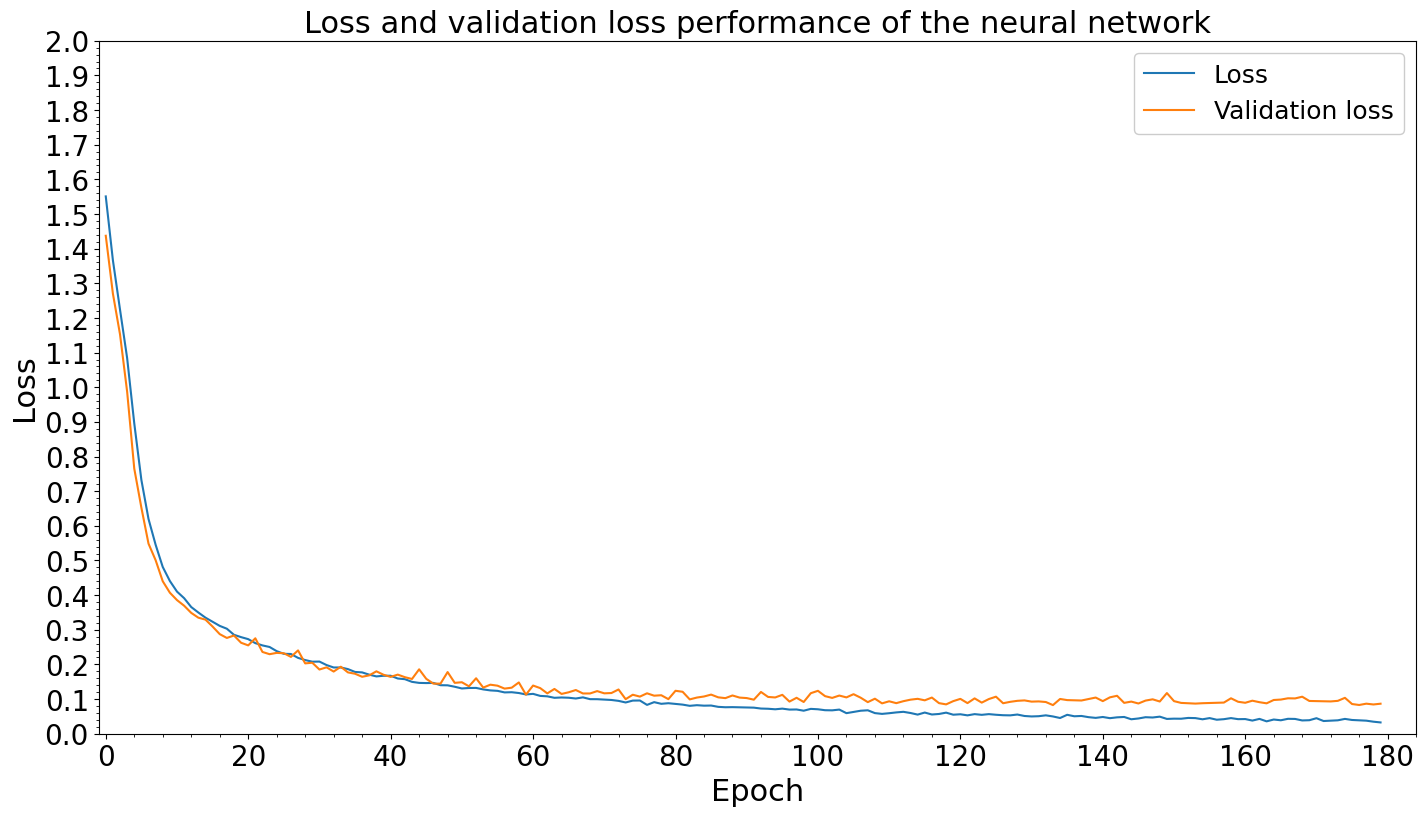}
     \caption{History curves of the neural network show how the
    training and validation loss evolved during the training and validation phases.}

    \label{fig:loss_performance}
\end{figure}

Figure\,\ref{fig:accuracy_performance} shows how both the training and validation accuracy evolved during the training and validation phase. Related to this, Figure\,\ref{fig:loss_performance} shows in detail, how the loss changed during this phase. For the training and testing process, we used a GPU-accelerated computer containing NVidia GeForce RTX 2080 Ti GPU cards. The TensorFlow environment we use is fully GPU-supported, therefore the computation benefits fully from the available GPUs.\\
The training phase took about 170 epochs, where one epoch lasted for 6 seconds. The whole training and validation phase took about 17 minutes. For training we used 14\,000 stars and for validation 6\,000 stars were included both in $g$ and $r$ filters.

Examination of the curves shows that the neural network performed well, beside some subtle anomalies the loss curves decreased and the accuracy curves increased continuously. By the 170th epoch the early stopping terminated the learning. As the training and validation metrics evolved in the same pace, the network did not overfit.

\subsection{Testing performance, classification of lightcurves}
\label{sec:testing_performance}

\begin{figure}
   \centering
   \includegraphics[width=\columnwidth]{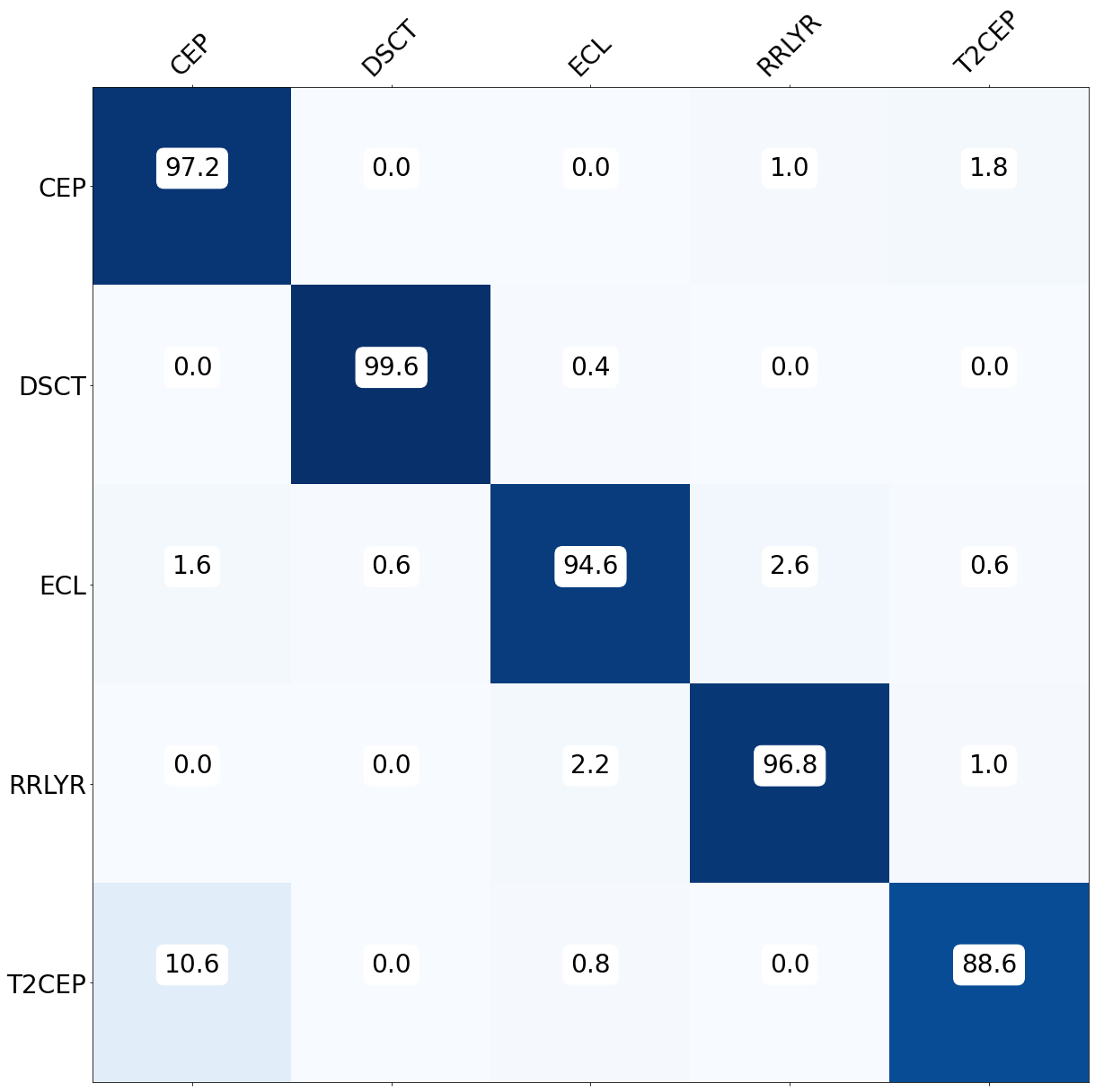}
     \caption{Confusion matrix of the testing phase, which shows the performance of our neural network using ZTF data. The inputs were phase-folded light curves in two ZTF passbands (\textit{g} and \textit{r}) and the given stars' period.}
    
    \label{fig:confusion_matrix}
\end{figure}

After the training and validation phase finished, we used the carefully selected test dataset and started the testing phase. The classification of the complete test dataset (2500 variable stars) took $0.7$ seconds. Figure\,\ref{fig:confusion_matrix} shows the result of the process. In the confusion matrix it is clearly visible that the classification accuracy is above $94$\% for four variable star classes -- for three classes it is above 98\% --, only Type II Cepheids have a bit lower classification accuracy result, but it is still around 90\%. The meaning of these definitions were detailed in our previous paper, see \citep{Szklenaretal2020}. 

The results of two related groups - Classical Cepheids and Type-II Cepheids - are worth looking at in more detail. The other three types, the $\delta$ Scuti stars, the eclipsing binaries and the RR Lyrae stars have almost negligible classification error.
For the Classical Cepheids $5.2$\% of the test stars were misclassified by the neural network, the majority of which - $4.4$\%  -are classified as Type-II Cepheids.

The classification result is worse for the Type-II Cepheids, $8.4$\% of these stars were identified as Classical Cepheid and $2$\% as eclipsing binary. 

As the classification error was slightly higher for these two types, we selected the stars that were misclassified from the tested stars and checked their light curves manually. It turned out, the problem is with the precision of the classification, not with the labels.

\subsection{Results using only one passband}
\label{sec:one_filter_only}

\begin{figure*}[h!]
   \centering
   \includegraphics[width=\textwidth]{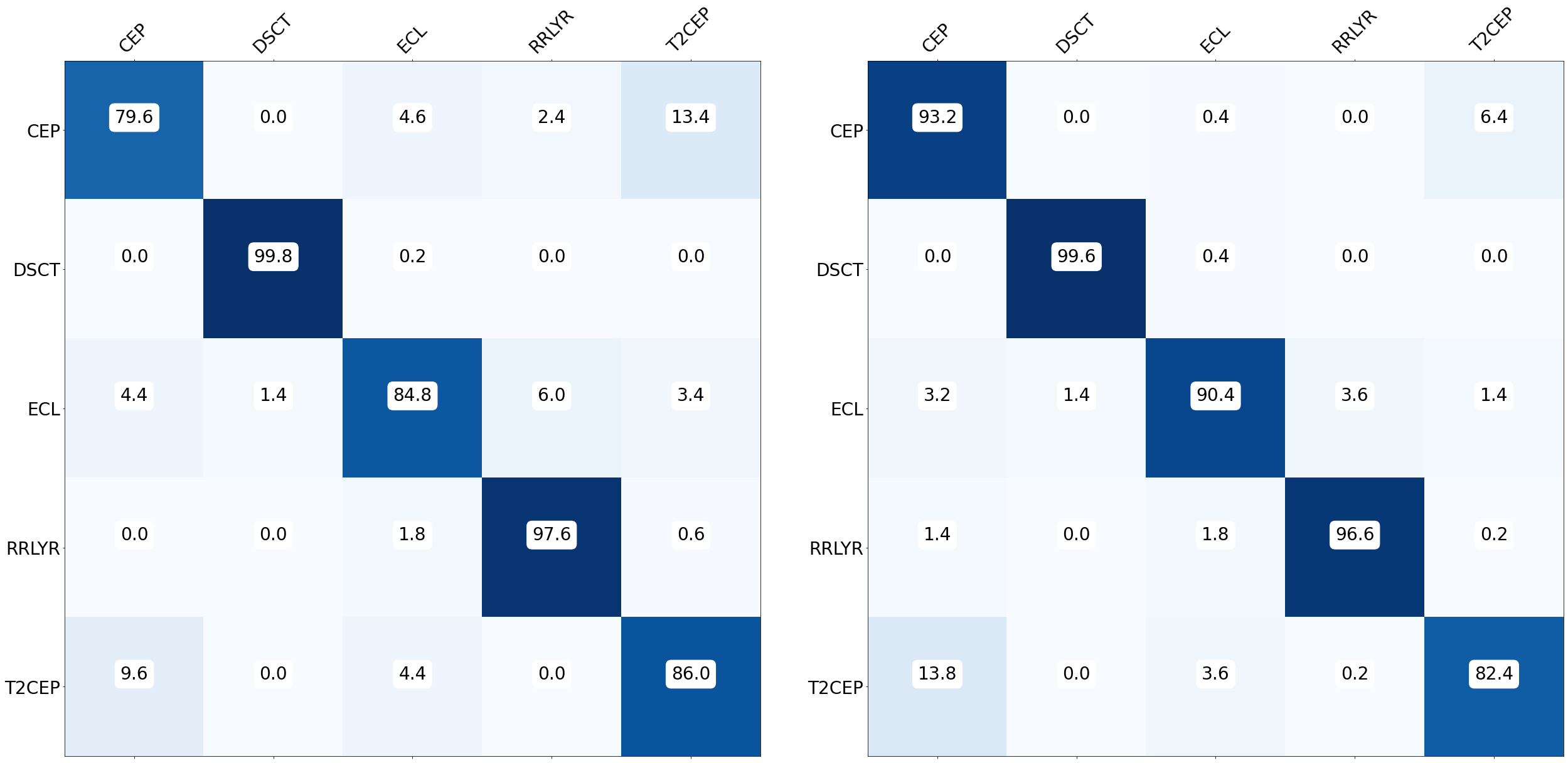}
     \caption{Test results using only one passband. The $g$ filter data are presented on the left and the $r$ filter data are on the right. Periods were used as additional numerical inputs. The accuracy of the latter is close to those obtained with multiple passbands, but the accuracy of the classification of Cepheids is significantly worse.
     }
    
    \label{fig:only_one_passband}
\end{figure*}

As our previously published works have shown, the classification of periodic variable stars with only one passband observations can achieve accurate results, even more so with additional numerical data (like periods). We ran single filter tests with the neural network to compare the performance with multiple passband results. Figure \ref{fig:only_one_passband} shows the classification result of the test sample when only one filter is used. Similarly to the architecture of the multiple passband neural network, in this case the phase-folded light curve images and the periods as numerical data were used as inputs. Comparing this figure to the previous results shows that it is worth combining data from different passbands to improve the accuracy of the most dubious classes.

\section{Summary and conclusions}
\label{sec:conclusions}

{In this paper, we have trained and tested a neural network that can perform classification of light curves obtained in a multiple color filters using their associated period as numerical data.
During the pre-processing, we selected variable stars from the data collected by the ZTF sky survey program that are also included in the OGLE-IV database. This choice was made because the latter contains a well-categorized, reliably classified data set. Since ZTF data are similar to the upcoming LSST observations in sampling, sky coverage and passbands, they demonstrate us how and with what efficiency we can classify periodic variable stars observed in multiple filters. 

In order to cross-match OGLE-IV periodic variable stars with the public ZTF object catalog, we used the Gaia astrometric measurements. During this process we involved probabilistic comparison of the targets'
coordinates and brightness. When the Gaia RP magnitudes were available, we compared them with the OGLE-IV I magnitudes and ruled out every sources, which do not match our criteria. 

The ZTF observes in $g$, $r$, and $i$ filters, but the latter is only partially available in the public data, so it was omitted and only $g$ and $r$ filters were used. The ZTF website \footnote{ZTF website: https://www.ztf.caltech.edu/index.html} provides a multiple object search feature, where the coordinates were given as input and we received a filtered list of sources based on the nearest neighbour method. This was further filtered based on the previously matched Gaia coordinates and brightness data.

Since the chosen different variable star classes, Classical Cepheids, $\delta$ Scuti stars, eclipsing binaries, RR Lyrae stars and Type II Cepheids, are represented in vastly different numbers in our sample, it was necessary to create synthetic data. After the data augmentation process, we had 4500 light curves of each variably type in both $g$ and $r$ filters for both training and testing purposes.

For the classification of variable stars observed by the LSST, it is essential to know the period of the source, so we decided to determine this numerical parameter from the ZTF data, even though the period was known from OGLE. In addition to calculating the period, we ran statistics to see after how many years we would be able to calculate the period of different types of variable stars. Since both the ZTF and the future LSST will have very few data points per year and per band, for some variable types (for example multiperiodic delta Scuti stars, double or triple mode Cepheids or long period Mira semiregular variable stars) it will be particularly difficult not only to determine the period, but also to generate phase-folded light curves at the beginning of the 10-year survey. In case of the newly discovered variable stars our statistics show that we need at least three years of data  as a minimum to classify most of the pulsating and eclipsing variable stars. However, the situation will be much better for the periodic variables discovered previously, especially those with a securely determined period value.

Before training a neural network using multiple color filters, we also constructed a neural network to classify light curves with only one filter and its associated periods to see how the efficiency changes when only one filter is available. Our preliminary results showed an accuracy of 73-100 \% with the $g$ filter and 87-99 \% with the $r$ filter.

In contrast, the neural network that used the two filters together showed a classification probability of between 90-100 \%, that can be considered a significant improvement in the neural network training process and the classification accuracy compared to the one-filter approach.

When we started this project the actual ZTF data release was DR15, from which our dataset was upgraded to DR17. As almost 2 years passed since than, the latest release is DR23 which contains much more data. Our tests will continue with these and even newer ones, hopefully we can extend the research to 3 different filters (g,r and i) as well. This would be very useful before the start of LSST, which will observe in 6 different filters (ugrizY).

Processing the large amount of data collected by the LSST using $6$ filters poses a huge challenge. The quality of the data generated during the 10-year long sky survey program is well demonstrated by the LSST Data Preview 1 (DP1) data package released after the first submission of our paper. The DP1 cadence in selected sky areas mimics the 10-year LSST survey cadence \citep{Malanchev_2025arXiv250623955M}. Indeed, these authors publish $ugrizy$ light curves of several known variable (pulsating and eclipsing) stars. The next step in our research will be also towards these directions to benefit from multiband LSST observations by using multiband period search methods, and by developing corresponding neural networks.

\section*{Data availability}
The architecture of the neural network and the output weight file can be downloaded from our research team's website, using the following link: \\ \url{https://konkoly.hu/KIK/data_en.html#ML}\\

\begin{acknowledgements}

This project is funded by the SNN-147362 grant  of  the  Hungarian  National Research,  Development  and  Innovation  Office. This project is also supported by the  SeismoLab project which is funded by the KKP-137523 grant and the Élvonal (Forefront) Research Excellence Program of the  National Research, Development and Innovation Office.
On behalf of the \texttt{Varclass –– Variable star classification} project we are grateful for the possibility to use HUN-REN Cloud (see \citep{HUN-REN_Cloud_H_der_2022}; https://science-cloud.hu/) which helped us to achieve the results published in this paper.
This work has made use of data from the European Space Agency (ESA) mission {\it Gaia} (\url{https://www.cosmos.esa.int/gaia}), processed by the {\it Gaia} Data Processing and Analysis Consortium (DPAC,
\url{https://www.cosmos.esa.int/web/gaia/dpac/consortium}). Funding for the DPAC has been provided by national institutions, in particular the institutions participating in the \texttt{Gaia} Multilateral Agreement.
Based on observations obtained with the Samuel Oschin Telescope 48-inch and the 60-inch Telescope at the Palomar Observatory as part of the Zwicky Transient Facility project. ZTF is supported by the National Science Foundation under Grants No. AST-1440341 and AST-2034437 and a collaboration including current partners Caltech, IPAC, the Weizmann Institute for Science, the Oskar Klein Center at Stockholm University, the University of Maryland, Deutsches Elektronen-Synchrotron and Humboldt University, the TANGO Consortium of Taiwan, the University of Wisconsin at Milwaukee, Trinity College Dublin, Lawrence Livermore National Laboratories, IN2P3, University of Warwick, Ruhr University Bochum, Northwestern University and former partners the University of Washington, Los Alamos National Laboratories, and Lawrence Berkeley National Laboratories. Operations are conducted by COO, IPAC, and UW.
Last but not least, we would like to thank our colleague, László Molnár, for his professional advice, which contributed to the completion of our article.
\end{acknowledgements}

%
%


\bibliographystyle{aa} 
\bibliography{references}{}

\appendix

\section{Randomized phase test}

\begin{figure}
   \centering
   \includegraphics[width=\columnwidth]{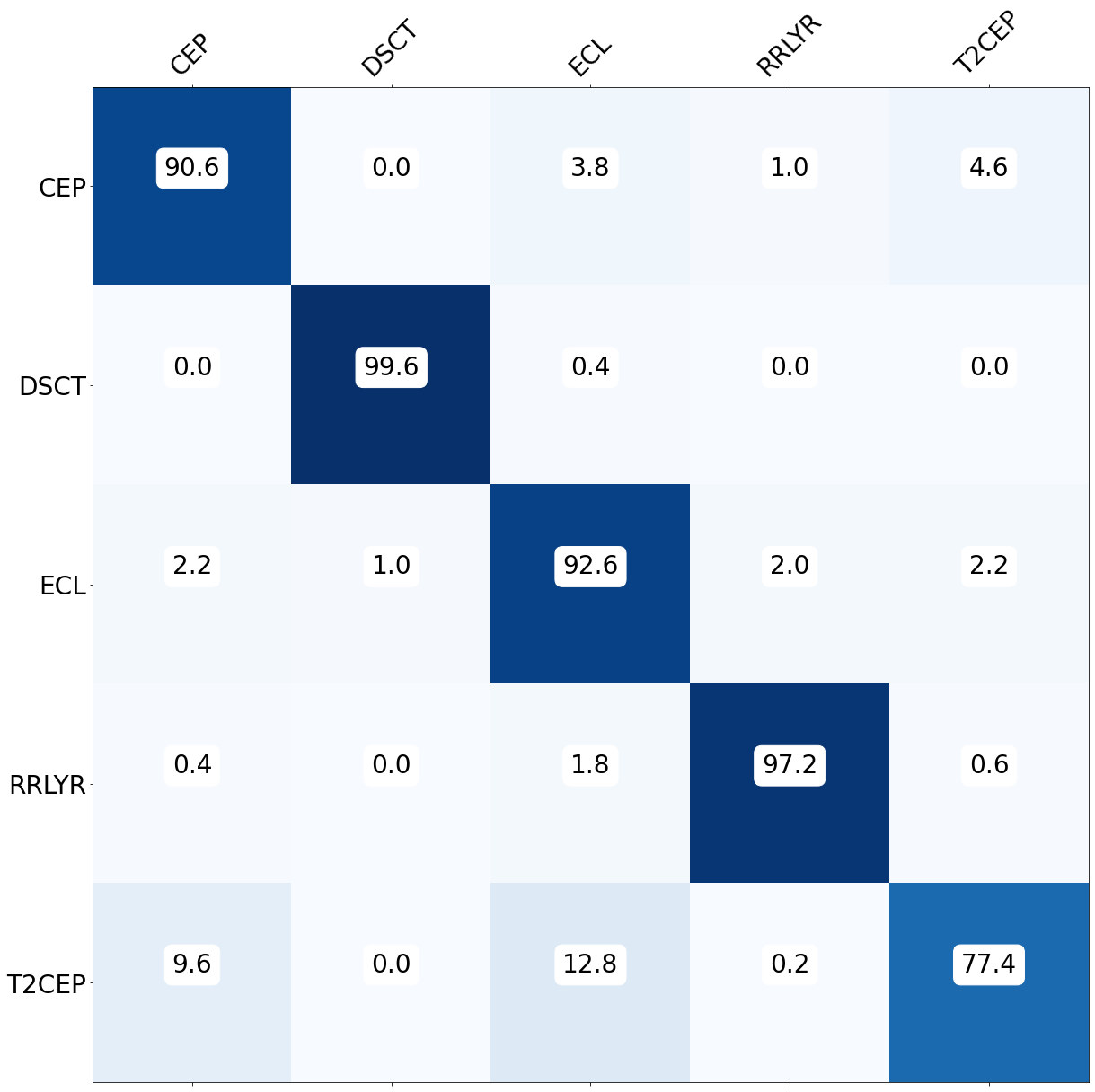}
     \caption{At the referee's request, we performed a test how our neural network architecture would perform with randomized phase. For this test we created a completely new training and testing dataset where the phase of the maximum/minimum value in the 0-1 interval was randomized. In this case the light curve minima of the eclipsing binaries and the maxima of the pulsating variable stars should not have any external information. Our results show that our neural network does not learn the phase information, only the light curve shape.}
    
    \label{fig:appendix_randomphase_cm}
\end{figure}

\appendix

\end{document}